\documentclass[twoside]{article}
\usepackage{graphicx} 
\usepackage{xcolor}
\usepackage{tikz}
\usepackage{hyperref}
\usepackage{soul}
\usepackage{stackengine}
\usepackage{xparse}
\usepackage{tokcycle}
\usepackage{xstring}
\usepackage{pgffor}
\usepackage{setspace}
\usetikzlibrary{plotmarks}
\usepackage{pgfplots}
\usepackage{pgfplotstable}
\pgfplotsset{compat=1.18}
\usetikzlibrary{patterns}
\usepackage{tikzpagenodes}
\usetikzlibrary{backgrounds}
\usetikzlibrary{positioning}
\usetikzlibrary{calc}
\usepackage{subcaption}
\usepackage{xparse}
\usepackage{amsmath}
\usepackage{changepage}

\usepackage[landscape,twocolumn]{geometry}
\usepackage{fancyhdr}
\makeatletter
\newcommand{\leftcolx}{
  \dimexpr \oddsidemargin + 1in + (\textwidth - \columnsep)/4\relax
}

\makeatother

\hypersetup{
    colorlinks,
    linkcolor={gray},
    citecolor={gray},
    urlcolor={gray}
}

\setul{0.3ex}{1.5pt}

\usepackage[backend=biber, style=numeric-comp, isbn=false, doi=true, url=false, eprint=false, maxnames=99]{biblatex}
\AtEveryBibitem{%
  \clearfield{note}%
}

\usepackage[T1]{fontenc}
\usepackage{lipsum}

\newlength\dellet

\newcommand{\setblotseed}[1]{%
  \pgfmathsetseed{#1}%
}
\setblotseed{3}

\pgfmathsetseed{1236}
\def\nVariants{3}

\newcommand{\drawRandomMargins}{%
  \pgfmathtruncatemacro{\randLraw}{3 + int(\nVariants * rand)}
  \pgfmathtruncatemacro{\randRraw}{3 + int(\nVariants * rand)}

  \begin{tikzpicture}[remember picture, overlay]
    \node[anchor=south west] at (current page.south west) 
      {\includegraphics[height=\paperheight]{scanline\number\randLraw.png}};

    \begin{pgfonlayer}{background}
      \node[anchor=south] at (current page.south) 
        {\includegraphics[height=\paperheight]{centerline\number\randLraw.png}};
    \end{pgfonlayer}
    
    \node[anchor=south east] at (current page.south east) 
      {\includegraphics[height=\paperheight]{scanline\number\randRraw.png}};
  \end{tikzpicture}%
}

\usepackage{atbegshi}

\AtBeginShipout{%
  \pgfmathtruncatemacro{\angl}{0 + int(2 * rand)}
  \setbox\AtBeginShipoutBox=\hbox{%
    \rotatebox{\angl}{\drawRandomMargins\box\AtBeginShipoutBox}%
  }%
}

\title{\textbf{{Investigating the Task Load} \\ {of Investigating the Task Load} \\ in {Visualization Studies}}}
\author{{Daniel Pahr\textsuperscript{\dag} and Sara Di Bartolomeo\textsuperscript{\ddag}}}
\date{}

\newcommand{\tlxtlx}{$\text{TLX}_{\text{TLX}}$}
\newcommand{\tlxtask}{$\text{TLX}_{\text{task}}$}

\begin{document}

\definecolor{lgray}{HTML}{D9D9D9}

\newcommand{\drawTLXblock}[2]{%
  \begin{scope}[shift={(#1,#2)}]
    \draw [pattern=crosshatch dots, pattern color=lgray] (0,0) rectangle (3,4);

    \foreach \x in {0.5, 0.75, ..., 2.5}
      \draw (\x,3) -- (\x,3.2);
    \draw (0.5, 3) -- (2.5, 3);
    \node at (1.5, 3.5) {\footnotesize\textbf{EFFORT}};

    \foreach \x in {0.5, 0.75, ..., 2.5}
      \draw (\x,2) -- (\x,2.2);
    \draw (0.5, 2) -- (2.5, 2);
    \node at (1.5, 2.5) {\footnotesize\textbf{PERFORMANCE}};

    \foreach \x in {0.5, 0.75, ..., 2.5}
      \draw (\x,1) -- (\x,1.2);
    \draw (0.5, 1) -- (2.5, 1);
    \node at (1.5, 1.5) {\footnotesize\textbf{FRUSTRATION}};

    \node at (1.5, 0.5) {\footnotesize\textbf{[...]}};
  \end{scope}
}

\newcommand{\radioComparison}[4]{%
  \begin{scope}[shift={(#1,#2)}]
    \node[anchor=south west, font=\footnotesize] at (0,1.85) 
      {What is more};
    \node[anchor=south west, font=\footnotesize] at (0,1.5) 
      {important?};

    \draw (0.2,1.45) circle (0.08);
    \node[anchor=west, font=\footnotesize] at (0.3,1.45) {#3};

    \draw (0.2,1.1) circle (0.08);
    \node[anchor=west, font=\footnotesize] at (0.3,1.1) {#4};
  \end{scope}
}

\NewDocumentCommand{\nasaChart}{m m o}{%
  \begin{figure}
  \tikzset{every path/.style={line width=1.5pt}}
  \begin{tikzpicture}[x=0.8cm, y=0.04cm]

    \draw (0,0) -- (9,0); 
    \draw (0,0) -- (0,100);
    \foreach \y in {0,20,...,100} {
      \draw (-0.1,\y) -- (0,\y);
      \node[left, font=\bfseries] at (-0.2,\y) {\y};
    }
    \node[rotate=90, font=\bfseries] at (-1.3,50) {RATING};

    \IfValueT{#3}{%
      \foreach \y in {#3} {
        \draw[dashed] (-0.1,\y) -- (9,\y);
      }
    }

    \def\xpos{0.2}

    #1

    \pgfmathsetmacro{\midpoint}{\xpos/2}

  \end{tikzpicture}
  \caption{#2}
  \end{figure}
}

\NewDocumentCommand{\nasaChartHorizontal}{m m o}{%
  \tikzset{every path/.style={line width=1.5pt}}
  \begin{tikzpicture}[x=0.05cm, y=0.55cm]

    \draw (0,0) -- (100,0); 
    \draw (0,0) -- (0,8);
    \foreach \x in {0,20,...,100} {
      \draw (\x,-0.1) -- (\x,0);
      \node[below, font=\bfseries] at (\x,-0.2) {\x};
    }
    \node[font=\bfseries] at (50,-1.5) {RATING};

    \IfValueT{#3}{%
      \StrBefore{#3}{;}[\mainlines]%
      \StrBehind{#3}{;}[\speciallines]%
      \foreach \x in \mainlines {
        \draw[dashed] (\x,-0.1) -- (\x,8);
      }
      \foreach \x in \speciallines {
        \draw[dash pattern=on 0pt off 2pt, line cap=round, gray] (\x,-0.1) -- (\x,8);
      }
    }

    \def\ypos{0.2}

    #1

  \end{tikzpicture}
}

\newcommand{\nasaWhiskerBar}[6]{%
  \pgfmathsetmacro{\xleft}{\xpos}
  \pgfmathsetmacro{\xright}{\xpos + #4}
  \pgfmathsetmacro{\xcenter}{\xpos + #4/2}
  \pgfmathsetmacro{\yvalue}{#2}
  \pgfmathsetmacro{\ymin}{\yvalue - #5}
  \pgfmathsetmacro{\ymax}{\yvalue + #6}
  \draw[pattern=#3] (\xleft,0) rectangle (\xright,\yvalue);
  \node[font=\bfseries] at (\xcenter, -5) {#1};
  \draw (\xcenter,\ymin) -- (\xcenter,\ymax);
  \draw 
    (\xcenter - 0.15,\ymin) -- (\xcenter + 0.15,\ymin)
    (\xcenter - 0.15,\ymax) -- (\xcenter + 0.15,\ymax);
  \pgfmathsetmacro{\xpos}{\xright + 0.2}
}

\newcommand{\nasaWhiskerBarHorizontal}[6]{%
  \pgfmathsetmacro{\ybottom}{\ypos}
  \pgfmathsetmacro{\ytop}{\ypos + #4}
  \pgfmathsetmacro{\ycenter}{\ypos + #4/2}
  \pgfmathsetmacro{\xvalue}{#2}
  \pgfmathsetmacro{\xmin}{\xvalue - #5}
  \pgfmathsetmacro{\xmax}{\xvalue + #6}
  \pgfmathsetmacro{\xright}{#4} 

  \draw[pattern=#3] (0,\ybottom) rectangle (#2,\ytop);

  \node[font=\bfseries, anchor=east] at (-0.2,\ycenter) {#1};

  \draw (\xmin,\ycenter) -- (\xmax,\ycenter);
  \draw 
    (\xmin,\ycenter - 0.15) -- (\xmin,\ycenter + 0.15)
    (\xmax,\ycenter - 0.15) -- (\xmax,\ycenter + 0.15);

  \pgfmathsetmacro{\ypos}{\ytop + 0.2}
}

\maketitle

\begin{figure}[b]
    \centering
    {\fontseries{m}\selectfont \textsuperscript{\dag}University of Vienna and \textsuperscript{\ddag}TU Wien, Vienna, Austria.}
     \\
     \hspace{1cm}
     \\
    This document was made to resemble the original NASA TLX Paper and Pencil Package (\url{https://humansystems.arc.nasa.gov/groups/tlx/downloads/TLX_pappen_manual.pdf}). 
\end{figure}

\newpage
\begin{abstract}
\begin{adjustwidth}{0.5cm}{0.5cm} 
    The NASA task load index (short: NASA-TLX) is a common metric to evaluate the workload of a user in a visualization study.
Yet, it is rarely performed as initially intended, as the sources-of-workload evaluation is often omitted for various reasons.
We conduct an online survey to investigate the task load of administering different versions of the NASA-TLX in a meta-study using the ReVISit framework.
Our results show that it is not the slight increase in experiment time, but rather participants' frustration with the procedure, that contributes to the slight increase in task load when using the full version of the TLX compared to using a shortened version.
However, we also show that the full version can shine a different and more faceted light on workload by adding a personal dimension to the data.
We propose that a compact version of the sources-of-workload questionnaire can mitigate both time loss and frustration for study participants, while still providing the same data as the original procedure.
The online study can be found and interactively explored on \url{https://dpahr.github.io/tlxtlx/}, and the source for the study, as well as the code for our analysis, can be found on \url{https://github.com/dpahr/tlxtlx/}.
\end{adjustwidth}
\end{abstract}
\section{Introduction}



A central element of many evaluations --- in the field of visualization and beyond --- is the measurement of the subjective task load on a user.
This can, for example, shine a light on the potential of a tool to lighten the cognitive load of users in comparison to other state-of-the-art methods.
Among all the tools in the human-factors toolbox, few have earned the kind of celebrity status that the NASA Task Load Index (NASA-TLX) enjoys.
Born in the cockpit but well established in usability labs, hospitals, control rooms, and even classrooms, NASA-TLX has become the de facto method for asking people one simple thing: How hard was that task, really?
This is done by rating six separate items on a questionnaire, dividing task load into its individual parts: mental, physical, and temporal demand, performance, effort, and frustration. 

The procedure for the NASA-TLX is well illustrated and disseminated via online resources, such as the paper and pencil package~\cite{hart_nasa_1986} or the computerized version~\cite{hart_nasa_1986-1}.
Nevertheless, \ul{the procedure is often modified}~\cite{hart_nasa-task_2006}, making it difficult to compare results between different publications and distorting the original meaning of this important metric. 
The most common modification is to omit performing a user-specific sources of workload evaluation, which provides weights for the individual dimensions of the TLX.\looseness=-1


We argue that, specifically for the analysis of visualizations using modern survey platforms, the complete procedure for the NASA-TLX is not only easy to implement but also quickly executed.
Thus, in this paper, we turn the tables. Instead of using NASA-TLX to evaluate a tool, we evaluate NASA-TLX itself. 
\ul{Our goal is to measure the impact of applying different versions of the NASA-TLX} in a crowdsourced online study.
We create a survey using the reVISit framework~\cite{ding_revisit_2023}, that confronts participants alternatingly with i) no scale weighting, ii) a compact scale rating procedure, and iii) a scale rating procedure that resembles the original paper version.

By measuring the impact of using the full version of the NASA-TLX, we demonstrate that this version offers more insights with a reasonable impact on study participants. Furthermore, we argue that this impact can be mitigated by using a compact version of the sources-of-workload evaluation.


\section{Background}

\begin{figure}
\setstretch{1}
\tikzset{every path/.style={line width=1.5pt}}
MENTAL DEMAND \\
\textcolor{white}{aaa} How mentally demanding was the task? 
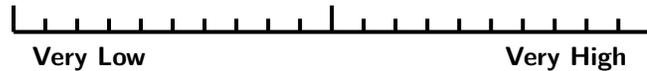
\begin{center}
\vspace{-0.6cm}
\begin{tikzpicture}[y=.2cm, x=0.05 * \linewidth, font=\sffamily]
\draw (0,0) -- coordinate (x axis mid) (20,0);

\foreach \x in {0}
\draw (\x,0pt) -- (\x,10pt)
node[anchor=south] {};

\foreach \x in {1,...,9}
\draw (\x,0pt) -- (\x,5pt)
node[anchor=south] {};

\foreach \x in {10}
\draw (\x,0pt) -- (\x,10pt)
node[anchor=south] {};

\foreach \x in {11,...,19}
\draw (\x,0pt) -- (\x,5pt)
node[anchor=south] {};

\foreach \x in {20}
\draw (\x,0pt) -- (\x,10pt)
node[anchor=south] {};
\end{tikzpicture}

\hspace{0.25cm} Very Low \hfill Very High \hspace{0.05cm}

\end{center}


PHYSICAL DEMAND \\
\textcolor{white}{aaa} How physically demanding was the task?
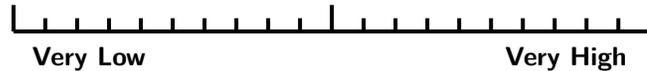
\begin{center}
\vspace{-0.6cm}
\begin{tikzpicture}[y=.2cm, x=0.05* \linewidth,font=\sffamily]
\draw (0,0) -- coordinate (x axis mid) (20,0);

\foreach \x in {0}
\draw (\x,0pt) -- (\x,10pt)
node[anchor=south] {};

\foreach \x in {1,...,9}
\draw (\x,0pt) -- (\x,5pt)
node[anchor=south] {};

\foreach \x in {10}
\draw (\x,0pt) -- (\x,10pt)
node[anchor=south] {};

\foreach \x in {11,...,19}
\draw (\x,0pt) -- (\x,5pt)
node[anchor=south] {};

\foreach \x in {20}
\draw (\x,0pt) -- (\x,10pt)
node[anchor=south] {};
\end{tikzpicture}

\hspace{0.25cm} Very Low \hfill Very High \hspace{0.05cm}

\end{center}


TEMPORAL DEMAND \\
\textcolor{white}{aaa} How hurried or rushed was the pace of the task?
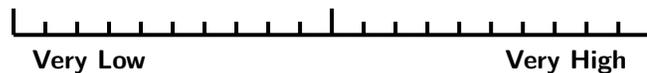
\begin{center}
\vspace{-0.6cm}
\begin{tikzpicture}[y=.2cm, x=0.05* \linewidth,font=\sffamily]
\draw (0,0) -- coordinate (x axis mid) (20,0);

\foreach \x in {0}
\draw (\x,0pt) -- (\x,10pt)
node[anchor=south] {};

\foreach \x in {1,...,9}
\draw (\x,0pt) -- (\x,5pt)
node[anchor=south] {};

\foreach \x in {10}
\draw (\x,0pt) -- (\x,10pt)
node[anchor=south] {};

\foreach \x in {11,...,19}
\draw (\x,0pt) -- (\x,5pt)
node[anchor=south] {};

\foreach \x in {20}
\draw (\x,0pt) -- (\x,10pt)
node[anchor=south] {};
\end{tikzpicture}

\hspace{0.25cm} Very Low \hfill Very High \hspace{0.05cm}

\end{center}


PERFORMANCE \\
\textcolor{white}{aaa} How successful were you in accomplishing what you were asked to do?
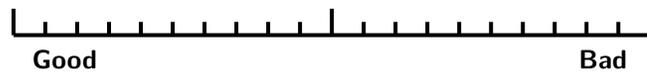
\begin{center}
\vspace{-0.6cm}
\begin{tikzpicture}[y=.2cm, x=0.05* \linewidth,font=\sffamily]
\draw (0,0) -- coordinate (x axis mid) (20,0);

\foreach \x in {0}
\draw (\x,0pt) -- (\x,10pt)
node[anchor=south] {};

\foreach \x in {1,...,9}
\draw (\x,0pt) -- (\x,5pt)
node[anchor=south] {};

\foreach \x in {10}
\draw (\x,0pt) -- (\x,10pt)
node[anchor=south] {};

\foreach \x in {11,...,19}
\draw (\x,0pt) -- (\x,5pt)
node[anchor=south] {};

\foreach \x in {20}
\draw (\x,0pt) -- (\x,10pt)
node[anchor=south] {};
\end{tikzpicture}

\hspace{0.25cm} Good \hfill Bad \hspace{0.05cm}

\end{center}


EFFORT \\
\textcolor{white}{aaa} How hard did you have to work to accomplish your level of performance?
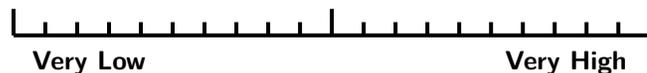
\begin{center}
\vspace{-0.6cm}
\begin{tikzpicture}[y=.2cm, x=0.05* \linewidth,font=\sffamily]
\draw (0,0) -- coordinate (x axis mid) (20,0);

\foreach \x in {0}
\draw (\x,0pt) -- (\x,10pt)
node[anchor=south] {};

\foreach \x in {1,...,9}
\draw (\x,0pt) -- (\x,5pt)
node[anchor=south] {};

\foreach \x in {10}
\draw (\x,0pt) -- (\x,10pt)
node[anchor=south] {};

\foreach \x in {11,...,19}
\draw (\x,0pt) -- (\x,5pt)
node[anchor=south] {};

\foreach \x in {20}
\draw (\x,0pt) -- (\x,10pt)
node[anchor=south] {};
\end{tikzpicture}

\hspace{0.25cm} Very Low \hfill Very High \hspace{0.05cm}

\end{center}


FRUSTRATION \\
\textcolor{white}{aaa} How insecure, discouraged, irritated, stressed, and annoyed were you?
\begin{center}
\vspace{-0.6cm}
\begin{tikzpicture}[y=.2cm, x=0.05* \linewidth,font=\sffamily]
\draw (0,0) -- coordinate (x axis mid) (20,0);

\foreach \x in {0}
\draw (\x,0pt) -- (\x,10pt)
node[anchor=south] {};

\foreach \x in {1,...,9}
\draw (\x,0pt) -- (\x,5pt)
node[anchor=south] {};

\foreach \x in {10}
\draw (\x,0pt) -- (\x,10pt)
node[anchor=south] {};

\foreach \x in {11,...,19}
\draw (\x,0pt) -- (\x,5pt)
node[anchor=south] {};

\foreach \x in {20}
\draw (\x,0pt) -- (\x,10pt)
node[anchor=south] {};
\end{tikzpicture}

\hspace{0.25cm} Very Low \hfill Very High \hspace{0.05cm}

\end{center}
\caption{A NASA-TLX questionnaire, as proposed by the paper and pencil package~\cite{hart_nasa_1986}.}
\label{fig:rtlx}
\end{figure}
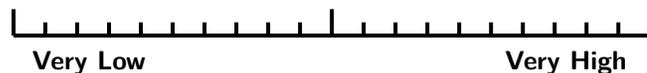

The NASA-TLX, developed by Hart and Staveland~\cite{hart_development_1988} in 1988, breaks subjective workload down into six intuitive components: mental demand, physical demand, temporal demand, effort, performance, and frustration. 
This can be done by marking a rating on a piece of paper or a PC, on a questionnaire similar to Figure~\ref{fig:rtlx}.
A slow interface might lead to user frustration, while a complicated task may lead to a higher cognitive load.
As task load is a highly subjective measure, each dimension is given a different weight for every individual, which can also vary across different tasks.
For example, one might care very little about temporal demand when there is no time limit for a task, or disregard physical effort when simply interacting with a computer using a mouse and keyboard.
\ul{The TLX is designed to account for participant-specific sources of workload}.
For each type of task, a participant is presented with all 15 possible pairings of the six scales in a random order, and for each, selects the one contributing higher to their subjective task load.
The participant then rates their experience in each of the categories on a scale from 0 to 100, and the final score --- the task load index --- is computed as a weighted average of the individual scores.


However, the procedure to acquire user-specific weights for the individual components is often disregarded.
This has sometimes been justified through a need for efficiency, since the weighting procedure takes time \cite{moroney_comparison_1992}.
Some research has been done to prove it to be redundant, due to a \ul{high correlation of the weighted and non-weighted TLX in some cases}~\cite{moroney_comparison_1992, byers_traditional_1989, hendy_measuring_1993}.
The abridged version without collecting individual weights for the scales has been called the ``raw'' task load index (RTLX).
Hendy et al.~\cite{hendy_measuring_1993} suggest that even a univariate rating, i.e., using a single scale for task load, would provide a good estimate of individual workload. 
We argue that \ul{using the complete procedure over the RTLX can provide additional insights that are otherwise lost}. While raw ratings may be used to pinpoint specific points of interest in an analysis, a raw average may vary greatly from the weighted average in specific cases~\cite{hart_development_1988}.

\ul{Multiple versions of the TLX questionnaire are available} directly from NASA, with precise instructions on how to use them.
The paper and pencil package~\cite{hart_nasa_1986} provides printable resources for a physical version, a computerized version instead proposes filling the questionnaires on a PC~\cite{hart_nasa_1986-1}, even an iOS version is provided~\cite{noauthor_tlx_nodate}.
Noyes and Bruneau~\cite{noyes_self-analysis_2007} compare the original paper and pencil package and the computerized version in a meta study, investigating if either of these versions incurs a higher task load on their study participants. They found the two versions to incur comparable task load.
Our study, while also a TLX metastudy where the TLX is applied as a metric to evaluate itself, is targeted to \ul{investigate different versions of the virtual questionnaire}.

Kosch et al.~\cite{kosch_survey_2023} warn of the ``hidden cost of the NASA-TLX''.
They argue that the TLX was not developed with HCI in mind, and that a cumulated score may obscure individual factors contributing to workload.
Still, they argue that its simplicity and ease of use are key advantages over other metrics.
In this paper, we seek to disentangle simplicity and malpractice using the NASA-TLX. 
We argue that using the RTLX brings no significant advantage over performing the full procedure.
The slightly longer procedure can be administered with \ul{little impact on study times and user experience}, using modern survey platforms.


\section{Experimental Setup}

\newcommand{\FHZamongUs}[3][]{%

      \draw[fill=#2, rounded corners = 3mm] (1.5,0) -- (1.5,5)
      arc (124.8074:103.8454:5) arc (80.6307:58.1808:5)-- (5,0) -- (3.7,0)
      {[rounded corners = 0mm] -- (3.7,1) coordinate(A) -- (2.7,1)}
      -- (2.7,0) -- cycle;
      \ifthenelse{\equal{#1}{angry}}
        {\draw[fill=#3]  plot[smooth cycle, tension=.7] coordinates
          {(4.3,4.7) (5.2,4.6) (5.2,3.5) (4.2,3.2) (3.1,3.5) (3.1,4.7)};}
        {\ifthenelse{\equal{#1}{very angry}}
          {\draw[fill=#3]  plot[smooth cycle, tension=.7] coordinates
            {(4.4,4.3) (5.2,4.6) (5.2,3.5) (4.2,3.6) (3.1,3.5) (3.1,4.7)};}
          {\draw[fill=#3]  plot[smooth cycle, tension=.7] coordinates
            {(4.3,4.9) (5.2,4.6) (5.2,3.5) (4.2,3.2) (3.1,3.5) (3.1,4.7)};}
        }

}

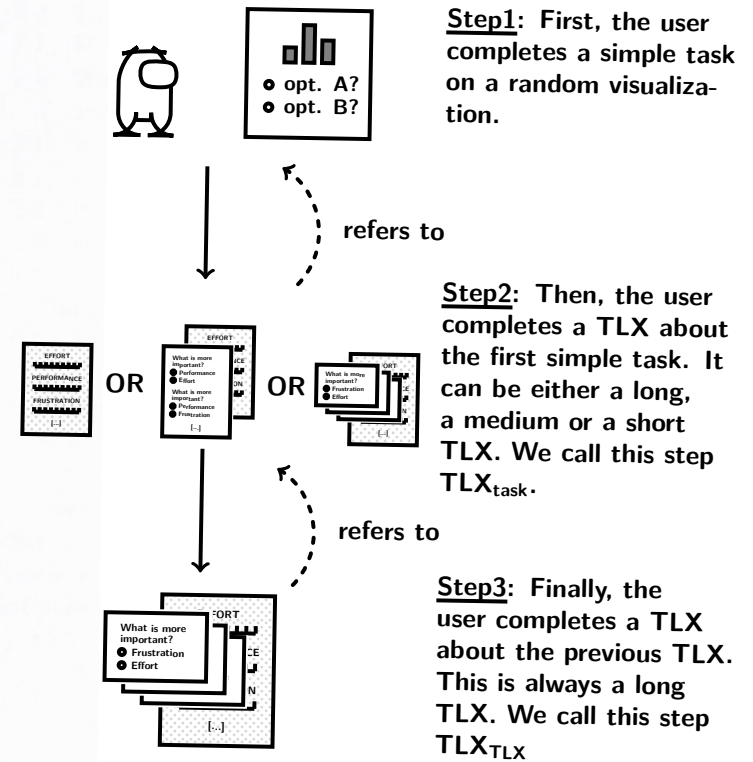
\begin{figure}[htbp]
    \centering

    \tikzset{every path/.style={line width=1.5pt}}

    \begin{tikzpicture}

    \coordinate (step1coord) at (0, 0);
    \coordinate (step2coord) at (0, -4);
    \coordinate (step3coord) at (0, -8);

    \node (step1) at (step1coord) {};
    \node (step2) at (step2coord) {};
    \node (step3) at (step3coord) {};
    
    \draw[->] ($(step1)+(0,-0.4)$) -- ($(step2)+(0,2.1)$);

    \draw[<-, dash pattern=on 2pt off 4pt, line cap=round]
        ($(step1)+(1,-0.4)$) arc (60:-50:1);
    \node at ($(step1)+(2.5,-1.2)$) {refers to};
    
    \draw[->] (step2) -- ($(step3)+(0,2.2)$);

    \draw[<-, dash pattern=on 2pt off 4pt, line cap=round]
        ($(step2)+(1,-0.4)$) arc (60:-50:1);
    \node at ($(step2)+(2.5,-1.2)$) {refers to};


    \begin{scope}[shift={($(step1)+(-1.5, 0)$)}, scale=0.2, transform shape]
        \FHZamongUs{white}{white}
    \end{scope}

    \begin{scope}[shift={($(step1)+(1,1)$)}, scale=0.5]
          \draw[fill=white] (-1, 1.4) rectangle (2.3, -2);
        
          \draw[fill=gray] (0,0) rectangle (0.3,0.4);
          \draw[fill=gray] (0.5,0) rectangle (0.8,1.0);
          \draw[fill=gray] (1,0) rectangle (1.3,0.6);
        
          \filldraw[fill=white] (-0.4, -0.6) circle (0.1);  
          \node [anchor=west, font=\footnotesize] at (-0.3, -0.6) {opt. A?};
          \filldraw[fill=white] (-0.4, -1.2) circle (0.1);   
          \node [anchor=west, font=\footnotesize] at (-0.3, -1.2) {opt. B?};

            
    \end{scope}

    \node[align=left, anchor=west, text width=4cm] at ($(step1coord)+(3,1)$)
    {\ul{Step1}: First, the user completes a simple task on a random visualization.};


    \begin{scope}[scale=0.3, transform shape, shift={($(step2coord)+(-8,0)$)}]
      \drawTLXblock{0}{0}
    \end{scope}

    \node at ($(step2coord)+(-1.05,0.7)$) {OR};

    \begin{scope}[scale=0.3, transform shape, shift={($(step2coord)+(-1.8,0)$)}]
        \drawTLXblock{1}{0.9}
        \draw [white, fill=white] (0,0) rectangle (3.05,4.05);
        \draw [fill=white] (0,0) rectangle (3,4);
        \radioComparison{0.25}{1.4}{Performance}{Effort}
        \radioComparison{0.25}{-0.1}{Performance}{Frustration}
        \node at (1.5, 0.4) {\footnotesize\textbf{[...]}};
    \end{scope}

    \node at ($(step2coord)+(+1.1,0.7)$) {OR};

    \begin{scope}[scale=0.3, transform shape, shift={($(step2coord)+(5.5,0)$)}]
        \drawTLXblock{1}{-.2}
        \draw [white, fill=white] (0.5,0.85) rectangle (3.25,2.7);
        \draw [fill=white] (0.5,0.9) rectangle (3.2,2.7);
        \radioComparison{0.75}{0.2}{Performance}{Effort}
        \draw [white, fill=white] (0,1.15) rectangle (2.75,3);
        \draw [fill=white] (0,1.2) rectangle (2.7,3); 
        \radioComparison{0.25}{0.5}{Performance}{Effort}
        \draw [white, fill=white] (-0.45, 1.45) rectangle (2.3,3.35);
        \draw [fill=white] (-0.5,1.5) rectangle (2.2,3.3); 
        \radioComparison{-0.25}{0.8}{Frustration}{Effort}
    \end{scope}

    \node[align=left, anchor=west, text width=4cm] at ($(step2coord)+(3,0.7)$)
    {\ul{Step2}: Then, the user completes a TLX about the first simple task. It can be either a long, a medium or a short TLX. We call this step \tlxtask.};


    \begin{scope}[scale=0.5, transform shape, shift={($(step3coord)+(-2,0)$)}]
        \drawTLXblock{1}{-.2}
        \draw [white, fill=white] (0.5,0.85) rectangle (3.25,2.7);
        \draw [fill=white] (0.5,0.9) rectangle (3.2,2.7);
        \radioComparison{0.75}{0.2}{Performance}{Effort}
        \draw [white, fill=white] (0,1.15) rectangle (2.75,3);
        \draw [fill=white] (0,1.2) rectangle (2.7,3); 
        \radioComparison{0.25}{0.5}{Performance}{Effort}
        \draw [white, fill=white] (-0.45, 1.45) rectangle (2.3,3.35);
        \draw [fill=white] (-0.5,1.5) rectangle (2.2,3.3); 
        \radioComparison{-0.25}{0.8}{Frustration}{Effort}
    \end{scope}

    \node[align=left, anchor=west, text width=4cm] at ($(step3coord)+(3,1)$)
    {\ul{Step3}: Finally, the user completes a TLX about the previous TLX. This is always a long TLX. We call this step \tlxtlx};
    
    \end{tikzpicture}
    
    \caption{Illustration of the experiment procedure in 3 steps.}
    \label{fig:experiment-steps}
\end{figure}

We implemented a small survey in reVISit~\cite{ding_revisit_2023}, comparing the impact of three different versions of the NASA-TLX on the task load of a study participant in a between-subjects design. 
The online study can be found and interactively explored on \href{https://dpahr.github.io/tlxtlx/}{our ReVISit instance}, and the source for the study, as well as the code for our analysis, can be found on \href{https://github.com/dpahr/tlxtlx/}{the GitHub repository}.

\subsection*{Procedure}

We show an illustration of the experiment procedure in Figure~\ref{fig:experiment-steps}.
Our study starts with an introduction for every participant. We give a brief overview of the procedure and make sure that participants are aware that they will complete a two-part experiment.
The first part comprises the participant performing a single visualization task from the mini-VLAT questionnaire~\cite{pandey_mini-vlat_2023} and being administered one of three versions of the NASA-TLX.
This \ul{first execution of the TLX is targeted to evaluate the participants' experience in completing the task}.

The second part has the participant completing another NASA-TLX questionnaire. 
This is always the full procedure, including the sources-of-workload evaluation, with the scales presented in pairs in a random order.
We carefully introduce the second part with a title card and a thorough introduction to make sure subjects understand that they are now \ul{evaluating the TLX procedure from part one}.

\subsection*{Conditions}

\begin{figure}
    \centering
    \tikzset{every path/.style={line width=1.5pt}}

    \makebox[.4\textwidth][c]{%
        \begin{subfigure}[b]{0.2\textwidth}
            \centering
            \begin{tikzpicture}  
                \drawTLXblock{0}{0}
            \end{tikzpicture}
            \label{cond:short}
            \caption{Condition (short): only the sliders are presented.}
        \end{subfigure}
        \hspace{1cm}
        \begin{subfigure}[b]{0.2\textwidth}
            \centering
            \begin{tikzpicture}    
                \drawTLXblock{1}{0.9}
                \draw [white, fill=white] (0,0) rectangle (3.05,4.05);
                \draw [fill=white] (0,0) rectangle (3,4);
                \radioComparison{0.25}{1.4}{Performance}{Effort}
                \radioComparison{0.25}{-0.1}{Performance}{Frustration}
                \node at (1.5, 0.4) {\footnotesize\textbf{[...]}};
            \end{tikzpicture}
            \label{cond:mid}
            \caption{Condition (medium): Sources-of-workload on a single page.}
        \end{subfigure}
    }

    \vspace{1cm}

    \begin{subfigure}[b]{0.4\textwidth}
        \centering
        \begin{tikzpicture}    
            \drawTLXblock{1}{-.2}
            \draw [white, fill=white] (0.5,0.85) rectangle (3.25,2.7);
            \draw [fill=white] (0.5,0.9) rectangle (3.2,2.7);
            \radioComparison{0.75}{0.2}{Performance}{Effort}
            \draw [white, fill=white] (0,1.15) rectangle (2.75,3);
            \draw [fill=white] (0,1.2) rectangle (2.7,3); 
            \radioComparison{0.25}{0.5}{Performance}{Effort}
            \draw [white, fill=white] (-0.45, 1.45) rectangle (2.3,3.35);
            \draw [fill=white] (-0.5,1.5) rectangle (2.2,3.3); 
            \radioComparison{-0.25}{0.8}{Frustration}{Effort}
        \end{tikzpicture}
            \label{cond:lonng}
        \caption{Condition (long): sources of workload on individual pages.}
    \end{subfigure}

    \caption{The three different conditions for the experiment. (a) Participants only complete the TLX questionnaire, no sources-of-worload evaluation (short). (b) Participants complete the sources-of-workload evaluation on a single page (medium). (c) Participants complete the sources-of-workload evaluation with successive comparisons, analogous to the paper and pencil package (long). }    
    \label{cond}
\end{figure}
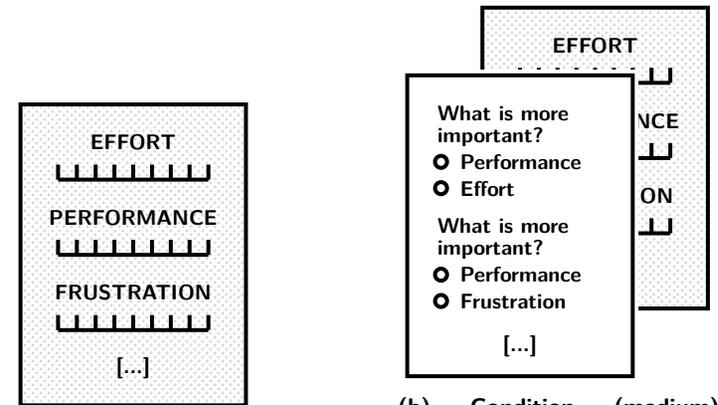
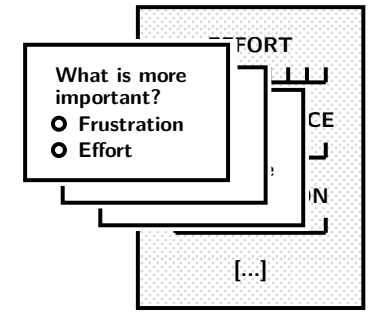

We compare three different versions of the NASA-TLX in our study (\textit{long, medium, short}). Figure~\ref{cond} shows an overview of our experiment conditions.
The \textit{long} condition represents the full TLX procedure, most closely resembling the original paper and pencil package. Participants are first shown \ul{pairs of scale names in a random order}, selecting the one that they deem to contribute more to their subjective task load. Afterwards, they are asked to rank their task load for the specific task on the six scales. The task load index is calculated as a weighted average of a user's ratings using the results of the sources-of-workload procedure.
As a \textit{medium} condition, with a predicted effort between the RTLX and the original version of the TLX, we use a compact form of the sources-of-workload evaluation. Instead of presenting the pairs of scales in succession, participants are shown a \ul{single page with all 15 pairs at once}. The task load is computed in the same way as in the \textit{long} condition.
In the \textit{small} condition, the NASA-TLX is administered in its abridged form, commonly known as the RTLX. The difference to the original version is that no sources-of-workload evaluation is performed, thus \ul{no weights are obtained} for the individual scales when computing the TLX score. The task load is computed as the average of a user's ratings. 

\subsection*{Metrics}
\sloppy

Our goal is to determine the impact of using the NASA-TLX on participants in a crowdsourced study.
The task the participants are asked to perform initially serves as a source of exertion, i.e., to create the workload we wish to evaluate.
Our participants complete two separate NASA-TLX questionnaires, one to determine the task load of the visualization task (\tlxtask) and the other to determine the task load of the first administered TLX questionnaire (\tlxtlx).
In the \textit{long} and \textit{medium} conditions, we look at the correlation between the weighted (\tlxtask) and unweighted (RTLX$_{\text{task}}$) average of the ratings for the visualization task. 
Additionally, we compare the results from the sources-of-workload evaluation (w$_{\text{TLX}_{\text{task}}}$) and the participants' questionnaire responses ($\text{RTLX}_{\text{task}}$) in rating the visualization task.
For the evaluation of the TLX procedure, we look at time ($\text{t}_{\text{TLX}_{\text{task}}}$) and task load index (\tlxtlx) of our participants across the three conditions (\textit{long, medium, short}).
The time for completing the questionnaire is measured from the participant having read the introduction, which either leads to the sources-of-workload evaluation or directly to the rating questionnaire, to the completion of the rating questionnaire.
We compute the TLX for a task as a weighted average of scores and weights per scale, and the RTLX as the average of scores on each scale.

\section{Results}

\begin{figure}
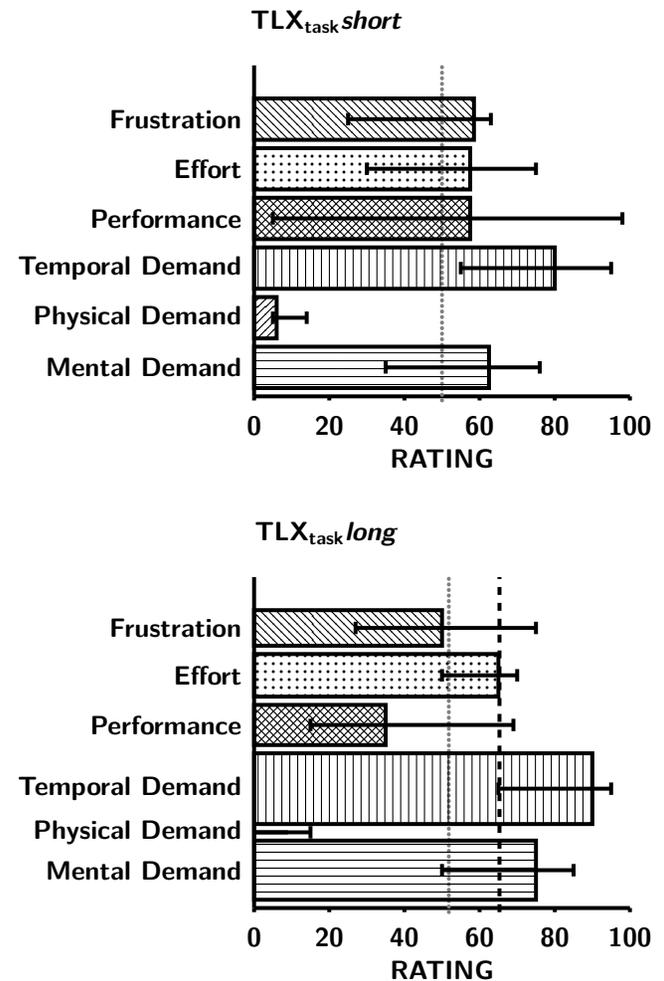

\centering

\tlxtask \textit{short} 

\hspace{0.5cm}

\nasaChartHorizontal{
    \nasaWhiskerBarHorizontal{Mental Demand}{62.5}{horizontal lines}{1}{27.5}{13.5}
    \nasaWhiskerBarHorizontal{Physical Demand}{6}{north east lines}{1}{1}{8}
    \nasaWhiskerBarHorizontal{Temporal Demand}{80}{vertical lines}{1}{25}{15}
    \nasaWhiskerBarHorizontal{Performance}{57.5}{crosshatch}{1}{52.5}{40.5}
    \nasaWhiskerBarHorizontal{Effort}{57.5}{dots}{1}{27.5}{17.5}
    \nasaWhiskerBarHorizontal{Frustration}{58.5}{north west lines}{1}{33.5}{4.5}
}{short (non-weighted, doesn't make sense on short) (tlx task)}[;50.00]

\tlxtask \textit{long}

\hspace{0.5cm}

\nasaChartHorizontal{
\nasaWhiskerBarHorizontal{Mental Demand}{75}{horizontal lines}{1.43}{25}{10}
\nasaWhiskerBarHorizontal{Physical Demand}{9}{north east lines}{0.00}{9}{6}
\nasaWhiskerBarHorizontal{Temporal Demand}{90}{vertical lines}{1.71}{25}{5}
\nasaWhiskerBarHorizontal{Performance}{35}{crosshatch}{0.97}{20}{34}
\nasaWhiskerBarHorizontal{Effort}{65}{dots}{1.03}{15}{5}
\nasaWhiskerBarHorizontal{Frustration}{50}{north west lines}{0.86}{23}{25}
}{}[65.29;51.81]

\caption{
The results for \tlxtask. A black dashed line indicates the average TLX score, while the grey dashed line indicates the unweighted average RTLX score. As we did not collect weights in the \textit{short} condition, only the RTLX score is shown.}
\label{res:tlxtask}
\end{figure}


We sent out an invitation to participate in our study via email to several visualization research groups.
We offered no reward for participation and did not collect demographic data. 
Of 34 responses in total, 20 completed the entire study, and we rejected 14 incomplete responses.
The number of samples received for each condition varied; we received seven responses for the \textit{long},  three responses for the \textit{medium}, and 10 responses for the \textit{short} version.
We focus mainly on the comparison of the \textit{long} and \textit{short} versions, and only briefly discuss implications of our findings on the \textit{medium} version due to the low number of responses in this condition.

We present the results similar to Hart~\cite{hart_nasa_1986} in their paper and pencil package.
They use a \ul{two-dimensional representation}, in the form of a bar chart, where the length of the bars represents the rating for the individual scale and the width of the bar represents the chosen weight.
We use this representation to show the median rating per scale and condition, and whiskers representing the quartiles. For the \textit{long} version, we display the average importance rating for each scale as the width of the bar. The overall average workload, i.e., the TLX, is shown as a dotted line across the chart.

\subsection*{Evaluation of a Visualization Task}

Figure~\ref{res:tlxtask} shows the results of the task load \ul{evaluation for the mini-VLAT task} that our participants completed.\looseness=-1

Comparing the RTLX of the \textit{long} version, i.e., calculating the raw average of scores instead of the weighted average, we see the average task load index roughly equal to the \textit{short} condition (51.81 vs 50). The RTLX is represented by the grey lines in Figure~\ref{res:tlxtask}, while the black lines represent the weighted TLX score.
Looking at the weighted average in the \textit{long} condition shows that the overall workload in the \textit{long} condition now much higher than in the \textit{short} condition (65.29 vs 50), while the ratings on the scales themselves are very similar.
We see an explanation for this in the average weighting.
Temporal and mental demand are weighed higher on average than frustration, effort, and performance, thus increasing the overall task load. On the other hand, physical demand was consistently rated unimportant by the participants, and does not have an impact on the TLX in the \textit{long} version, while in the \textit{short} version, it brings down the overall score. 

\subsection*{Evaluation of a TLX Questionnaire}

\begin{figure}
    \centering

    \tlxtlx \textit{short}

    \hspace{0.5cm}

    \nasaChartHorizontal{
    \nasaWhiskerBarHorizontal{Mental Demand}{47.5}{horizontal lines}{1.56}{17.5}{22.5}
    \nasaWhiskerBarHorizontal{Physical Demand}{4}{north east lines}{0.12}{4}{1}
    \nasaWhiskerBarHorizontal{Temporal Demand}{19.5}{vertical lines}{1.64}{19.5}{45.5}
    \nasaWhiskerBarHorizontal{Performance}{42.5}{crosshatch}{0.88}{27.5}{37.5}
    \nasaWhiskerBarHorizontal{Effort}{32.5}{dots}{0.76}{12.5}{27.5}
    \nasaWhiskerBarHorizontal{Frustration}{40}{north west lines}{1.04}{30}{25}
    }{}[43.10;]

    \tlxtlx \textit{long}

    \hspace{0.5cm}

    \nasaChartHorizontal{
    \nasaWhiskerBarHorizontal{Mental Demand}{55}{horizontal lines}{1.77}{24}{15}
    \nasaWhiskerBarHorizontal{Physical Demand}{5}{north east lines}{0.34}{5}{10}
    \nasaWhiskerBarHorizontal{Temporal Demand}{15}{vertical lines}{0.69}{12}{15}
    \nasaWhiskerBarHorizontal{Performance}{38}{crosshatch}{1.03}{38}{12}
    \nasaWhiskerBarHorizontal{Effort}{60}{dots}{1.03}{25}{0}
    \nasaWhiskerBarHorizontal{Frustration}{70}{north west lines}{1.14}{20}{20}
    }{}[52.43;]

\caption{Results for \tlxtlx. A black dashed line indicates the average TLX score.}
\label{res:tlxtlx}
\end{figure}

Figure~\ref{res:tlxtlx} shows the results of the task load \ul{evaluation for the task load index questionnaire} our participants completed. 

The median completion time for the TLX evaluation of the task was 81 seconds in the \textit{short} condition, which is almost twice as fast as in the \textit{long} condition with 156 seconds.
For the average task load, we see an increase in the \textit{long} version compared to the \textit{short} version (52.43 vs 43.10).
Interestingly, while we did measure that participants spent almost double the amount of time on the \textit{long} version than on the \textit{short} version (156 seconds vs 81 seconds), temporal demand in completing the task load questionnaire did not seem to impact our participants' task load much in either condition. 
The most apparent difference between the two conditions is the increased median frustration of the participants in the \textit{long} version compared to the \textit{short} version (70 vs 40), possibly caused by the high number of pairwise comparisons presented in succession.

\subsection*{The \textit{medium} Version}

\begin{figure}[h]
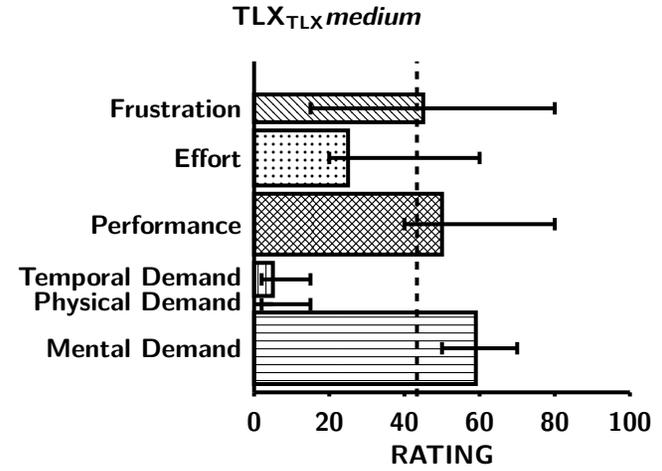

    \centering
    \tlxtlx \textit{medium}

    \hspace{0.5cm}

    \nasaChartHorizontal{
        \nasaWhiskerBarHorizontal{Mental Demand}{59}{horizontal lines}{1.73}{9}{11}
        \nasaWhiskerBarHorizontal{Physical Demand}{5}{north east lines}{0.00}{3}{10}
        \nasaWhiskerBarHorizontal{Temporal Demand}{5}{vertical lines}{0.80}{3}{10}
        \nasaWhiskerBarHorizontal{Performance}{50}{crosshatch}{1.47}{10}{30}
        \nasaWhiskerBarHorizontal{Effort}{25}{dots}{1.33}{5}{35}
        \nasaWhiskerBarHorizontal{Frustration}{45}{north west lines}{0.67}{30}{35}
    }{}[43.33;]
\caption{Results from \tlxtlx for the \textit{medium} condition. A black dashed line indicates the average TLX score.}
\label{res:tlxtlx:med}
\end{figure}

We only collected three samples for the \textit{medium} condition, hence we omit discussing the results of our analysis in comparison with the other two conditions. 
However, we still see it worthwhile to discuss the initial results in this separate section.

Figure~\ref{res:tlxtlx:med} shows the results for the experiment from the medium condition.
The average overall workload for the \textit{medium} condition is similar to the \textit{short} condition (43.33 vs 43.10). 
We observe again reduced temporal demand reported for this condition, even compared to the other two. 
The median rating for frustration in the \textit{medium} condition is also less than in the \textit{long} condition (45 vs 70), however the highest reported frustration score was 80 in the \textit{medium} condition.
The average completion time of the \textit{medium} questionnaire was 118 seconds, which is faster than the average (252 seconds) and median (156 seconds) of the \textit{long} version.  
While these results are promising for an initial observation, they should still be taken with a grain of salt, due to the low sample count.

\subsection*{Discussion}

At first sight, our study results seem to confirm the validity of criticism on the length of the full TLX procedure by Moroney et al.\cite{moroney_comparison_1992}. Our participants who completed a sources-of-workload evaluation spent more time than those who did not, leading to a longer total time participants spent on the experiment. 
However, our procedure saw participants complete only a single task in our experiment.
The sources-of-workload procedure is only completed once per task type; hence, \ul{for longer experiments, the additional time spent on the sources-of-workload evaluation also becomes relatively less of a contributor to the experiment time}. 

In their 20-year anniversary review~\cite{hart_nasa-task_2006} of the NASA-TLX, Sandra G. Hart, one of the researchers behind the TLX, reports: \begin{quote}
    In the 29 studies in which RTLX was compared to the original version, it was found to be either more sensitive (Hendy, Hamilton, \& Landry, 1993), less sensitive (Liu \& Wickens, 1994), or equally sensitive (Byers, Bittner, Hill, 1989), so it seems you can take your pick.
\end{quote}
We argue that \ul{the sources-of-workload procedure effectively measures a second dimension of task load}, which we try to indicate through encoding the average weights per scale into our figures.
Our representation shows personal preferences for a task by revealing the sources of workload of a study sample.

While we did not receive enough responses, we still observe an \ul{improvement in completion time and task when using the compact sources-of-workload evaluation} compared to the original version.
An advantage of the single-page version of the sources-of-workload evaluation may be that study participants can see how many pairs there are for them to rate, mitigating their frustration.

We received little written feedback in our survey, the only comment stating that the participant got confused with the scale names.
While we provided the descriptions as help text in ReVISit, we argue that \ul{users need consistent support in completing the TLX}, especially in crowd-sourced studies.
Another indicator for this is the high variance of the performance score in most of our experiments, which is counterintuitive in the TLX questionnaire since it ranges from ``good'' to ``poor'', while other scores all range from ``low'' to ``high''.
Here, we suggest including scale definitions directly on screen whenever referring to them, in any virtual setting.

Frequently, the TLX questionnaire is modified even beyond omitting the sources of workload evaluation.
In visualization studies, commonly performed in front of computer screens, using mainly mouse and keyboard, the physical component has been called into question or outright disregarded~\cite{victorelli2025we2}.
This may be sensible in isolated A/B comparisons; however, \ul{by modifying an established metric, the ability to compare results with prior studies is lost.} 
Beyond this, opting to report on individual scales instead of a compound score complicates statistical analysis, requiring adjustments in hypothesis testing such as Bonferroni correction. 
The TLX is designed to be a metric for task load; hence, if one desires to measure specific sub-items, then assumptions can only be made for that specific item, not for task load overall, and the reference to the TLX would be superfluous.

\subsection*{Conclusion}

Kosch et al.~\cite{kosch_survey_2023} speak of the ``hidden cost'' of the TLX. Instead, we want to highlight its \ul{forgotten benefits}.
Firstly, the NASA-TLX, while not initially intended for use in HCI, measures highly relevant dimensions of workload independent of the nature of the task.
Secondly, the personalized nature of the questionnaire allows us to analyze an otherwise hidden dimension of workload, lost to us if the sources-of-workload evaluation is not performed.
Finally, modern study frameworks allow us to execute a procedure that may have been cumbersome in the past with considerable ease. 
The NASA-TLX remains a useful measurement tool in our studies, now even for itself.\looseness=-1


\subsection*{Acknowledgement}

The authors thank \ul{Tingying He} from the University of Utah for providing us with the implementation of the NASA-TLX in ReVISit.

\subsection*{Conflict of Interest Statement}

Sara Di Bartolomeo is also involved in the organization of the workshop this document is submitted to, \href{https://altvis.github.io/}{alt.vis} 2025.

\newcommand{\nasaBar}[4]{%
  \pgfmathsetmacro{\xleft}{\xpos}
  \pgfmathsetmacro{\xright}{\xpos + #4}
  \pgfmathsetmacro{\xcenter}{\xpos + #4/2}
  \draw[pattern=#3] (\xleft,0) rectangle (\xright,#2);
  \node[font=\bfseries] at (\xcenter, -5) {#1};
  \pgfmathsetmacro{\xpos}{\xright + 0.2}
}

\section*{About the Look of this Document}

The look of the document is meant to emulate the appearance of the original NASA TLX Pen and Paper manual. What you are looking at is not a messy scan of a document --- instead, everything about this document is completely generated with \LaTeX, including scanlines and stapling in the middle of the booklet. All of the charts, drawings and diagrams are also just commands, generating everything using tikz. The source code for generating this document in \LaTeX will be distributed as part of the supplemental material.

\printbibliography

\end{document}